\documentclass[%
 aps,twocolumn,
 amsmath,amssymb,superscriptaddress, 
 prb,%
]{revtex4-1}

\usepackage{graphicx}
\usepackage{dcolumn}
\usepackage{bm}
\usepackage{color,soul}
\usepackage[usenames,dvipsnames]{xcolor}
\usepackage[colorlinks=true,linkcolor=RoyalBlue,citecolor=OliveGreen,urlcolor=OliveGreen,linktoc=page]{hyperref} 

\usepackage{enumerate}
\usepackage{grffile}
\usepackage{balance}
\usepackage{float}

\newcommand{\Hct}{H_\mathrm{c2}}

\newcommand{\Jdp}{J_\mathrm{dp}} 
\newcommand{\Jext}{J_{\mathrm{ext},x}}

\newcommand{\Tc}{T_\mathrm{c}}
\newcommand{\Tci}{T_\mathrm{c}^\star}
\newcommand{\ei}{\varepsilon^\star}

\begin{document}

\title{Dynamic Vortex-Mott Transition in 2D Superconducting Proximity Arrays}

\author{Andreas Glatz} 
\affiliation{Materials Science Division, Argonne National Laboratory, Argonne, Illinois 60439, USA}
\affiliation{Department of Physics, Northern Illinois University,  DeKalb, Illinois 60115, USA}
\author{Valerii Vinokur}
\affiliation{Materials Science Division, Argonne National Laboratory, Argonne, Illinois 60439, USA}

\date{\today}
\begin{abstract} 
The paradigmatic Mott insulator arises in strongly correlated systems, where strong local repulsion localizes interacting particles in underlying egg-holder-like potential.
The corresponding Mott transition reflects delocalization of the charges either by varying parameters of the system and temperature, or by applied current, the latter being referred to as the dynamic Mott transition. Recently, the dynamic Mott transition was experimentally observed on the vortex system trapped by the periodic proximity array and described in terms of non-Hermitian theory of nonequilibrium processes. Here we investigate numerically the vortex dynamic Mott transition in an proximity array implemented as an array of holes in a superconducting films and discover striking nonmonotonic behavior of the differential resistance as function of the applied current when deviating from the matching field corresponding to a unity filling factor.
\end{abstract} 

\maketitle

\section{\label{intro}Introduction}

Mott states arise in a system of strongly interacting repulsive particles placed on the underlying periodic lattice\cite{Mott:1949, Mott:1990}.
If at commensurate fillings the energy cost for delocalizing  
particles from the lattice sites is prohibitively high, the system falls into
a Mott insulating state. Increasing kinetic energy drives Mott insulator 
to a metallic state with itinerant particles\cite{Georges:1996,Balents:2005,Quantum:2012} via
Mott transition.
The transition is mediated by tunneling between the lattice cites and is driven by tuning 
of either the strength of correlations, or temperature, or underlying lattice potential.
If tunneling is promoted by the external driving field, the transition occurs at the out-of-equilibrium and is referred to as the dynamic Mott transition. 
As one of the central issues in the quantum physics of condensed matter, the nature of the Mott transition is a subject of intense research and has been supposedly seen in prototypical Mott systems like Cr-doped vanadium sesquioxide, (V$_{1-x}$Cr$_x$)$_2$O$_3$\,\cite{McWhan:1970,Jayaraman:1970,McWhan:1971,McWhan:1973,
	Kuwamoto:1980,Rosenberg:1995,Held:2001,Limelette:2003}, GaTa$_4$Se$_8$\cite{Camjayi:2014}, organic materials\cite{Furukawa}, and many others. Mott transition in two dimensional
lattice boson systems experienced explosive development with the advent of cold trapped atoms 
in an optical lattice, see\cite{Bloch:2008,Gemelke:2009} and references therein.
Yet, despite the extensive experimental work the detailed scenario for the Mott transition and its 
quantum critical nature remain not fully clear. The nature of \textit{dynamic} Mott transition 
is explored by far less.

\begin{figure}[htb]
	\includegraphics[width=0.6\columnwidth]{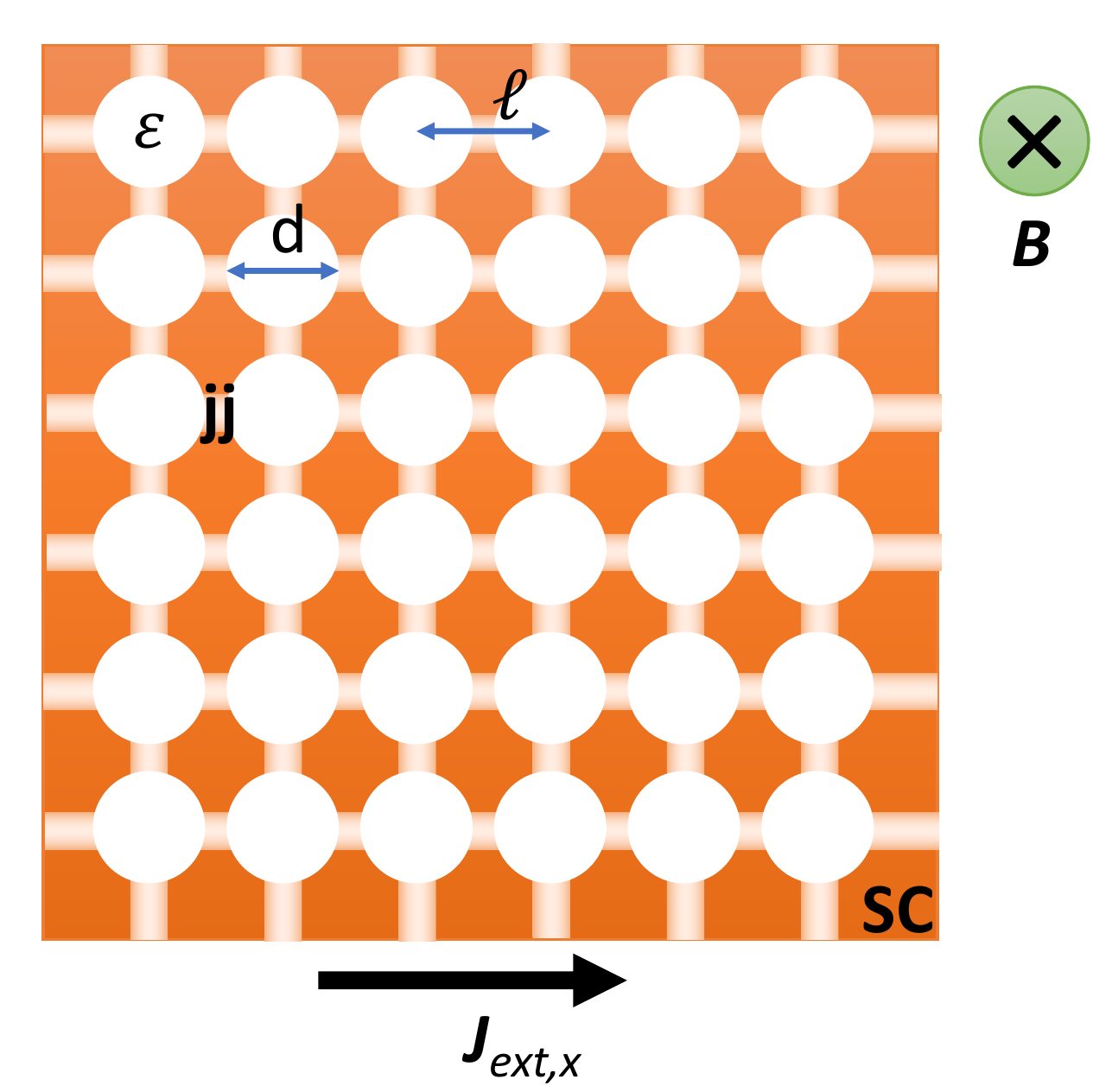}
	\caption{A two-dimensional (2D) superconducting film with a regular array of holes or metallic inclusions forming a 2D proximity array of superconducting (SC) islands. The lattice constant of the hole/inclusion array is given by $\ell$ and the hole diameter by $d$. The behavior of the inclusions is controlled by the parameter $\epsilon$, which is related to the critical temperature of the inclusion. A bias current $\Jext$ is applied in $x$-direction and the magnetic field $B$ perpendicular to the system. At narrow regions in the array, the order parameter is suppressed (indicated by the color gradient), forming a Josephson Junction (jj). The simulated system is periodic.}
	\label{fig:system}
\end{figure}

 In a recent experimental 
breakthrough\,\cite{Science:2015}, the current-driven Mott transition has been observed in a system of vortices pinned by a periodic array of 
proximity coupled superconducting islands.
Using vortex system in the proximity array enables to bypass detrimental effects of disorder inevitable in electronic systems. Varying magnetic field provides precise control over the vortex density. The ``Mottness'' is evidenced by critical behavior of the differential resistance at the commensurate vortex lattice fillings.
The revealed critical behavior of the dynamic resistance near the dynamic Mott critical point appeared to belong to the liquid-gas
transition universality class.  Remarkably, the dynamic vortex Mott transition has the same critical behavior as the thermodynamic electronic Mott transition up to the replacement of temperature by the applied current. Our findings enable the study of the Mott physics of strongly correlated quantum systems via classical vortices in superconducting arrays.  
The dynamic Mott criticality was successfully described by non-Hermitian theory of nonequilibrium processes\,\cite{Tripathi2016}. The possibility of this description suggests the topological nature of the Mott transition\,\cite{Galda2017}. This calls for careful study of the Mott dynamic behavior in a variety of systems. Here we explore an alternative proximity array designed as an array of holes in a superconducting films. Such a system has been extensively studied experimentally in connection with the studies of the superconductor-insulator transition, see\,\cite{Baturina2011,Postolova2017} for review.
 Here, we undertake a numerical study of the vortex Mott transition employing the time-dependent Ginzburg-Landau (TDGL) equation.
We reveale striking non-monotonic behavior of the differential resistance $dV/dI$ as function of the applied current at filling factors $f$ deviating from $f=1$, which differs the whole array from the proximity array obtained as an array of superconducting islands on the metallic substrate used in\,\cite{Science:2015}.

\section{Numerical Model}\label{sec:model_GL}

In order to model  superconducting patterned films, which constitutes the proximity array, we use the TDGL equation and apply it to systems with periodically modulated inclusion.
The TDGL equations effectively capture the collective vortex dynamics and pinning in realistic systems. 
As the London penetration depth is typically large compared to the coherence length in superconducting films, the TDGL equations can be simplified to just the time evolution of the superconducting order parameter, $\psi = \psi(\mathbf{r}, t)$, with a constant magnetic field $B$ perpendicular to the film, i.e.~\cite{Sadovskyy:2015a},
\begin{equation} 
(\partial_t + i \mu)\psi 
= \epsilon (\mathbf{r}) \psi - |\psi |^2 \psi + 
(\boldsymbol{\nabla} - i \mathbf{A})^2 \psi + \zeta(\mathbf{r}, t),
\label{eq:GL} 
\end{equation} 
where $\mu = \mu(\mathbf{r}, t)$ is the scalar potential, $\mathbf{A} = -yB\hat{\mathbf{e}}_x$ is the vector potential associated with the external magnetic field perpendicular to the film, and $\zeta(\mathbf{r}, t)$ is an additive thermal-noise term. This equation provides an adequate, quantitative description of strong {type-II} superconductors in the vortex phase. 
Equation~\eqref{eq:GL} is written in dimensionless units, where the unit of length is the superconducting coherence length~$\xi$, the unit of time is $t_0 \equiv 4\pi \sigma \lambda^2 / c^2$, $\lambda$ is the London penetration depth, $\sigma$ is the normal-state conductance, and the unit of the magnetic field is given by the upper critical field $\Hct = \hbar c / 2e \xi^2$ ($-e$ is the electron's charge and $c$ is the speed of light).
Thermal fluctuations, described by $\zeta(\mathbf{r}, t)$, are determined by its time and space correlations and absolute temperature.~\cite{Sadovskyy:2015a}

In order to study the dynamical response of the system, we apply a current to the system along the $x$-direction. 
The total (normal and superconducting) in-plane current density is then given by the expression 
\begin{equation}
\mathbf{J} 
= \mathrm{Im} \bigl[ \psi^*(\boldsymbol{\nabla} - i \mathbf{A}) \psi \bigr]
- \boldsymbol{\nabla} \mu,
\label{eq:J} 
\end{equation}
in units of $J_0 = \hbar c^2 / 8\pi e \lambda^2 \xi$. 
For an applied current density $\Jext$, we need to solve an additional ordinary differential equation for the voltage: $\Jext = \langle J_x \rangle_\mathbf{r}$, where $\langle \cdot \rangle_\mathbf{r}$ is the spatial average over the complete system. 
(The maximum theoretical depairing current density for a homogeneous system is $\Jdp = 2 J_0 / 3\sqrt{3}$.) 
Furthermore, we need to take current conservation, $\boldsymbol{\nabla} \mathbf{J} = 0$, into account, resulting in the Poisson equation for $\mu$:
\begin{equation}
\Delta\mu =
\boldsymbol{\nabla} \,
\mathrm{Im} [ \psi^*(\boldsymbol{\nabla} - i \mathbf{A}) \psi ]\,.
\label{eq:Poisson}
\end{equation}

This closed set of equations is solved on a high-performance GPU cluster for several million time steps.~\cite{Sadovskyy:2015a}

\begin{figure*}[htb]
	\includegraphics[width=0.8\textwidth]{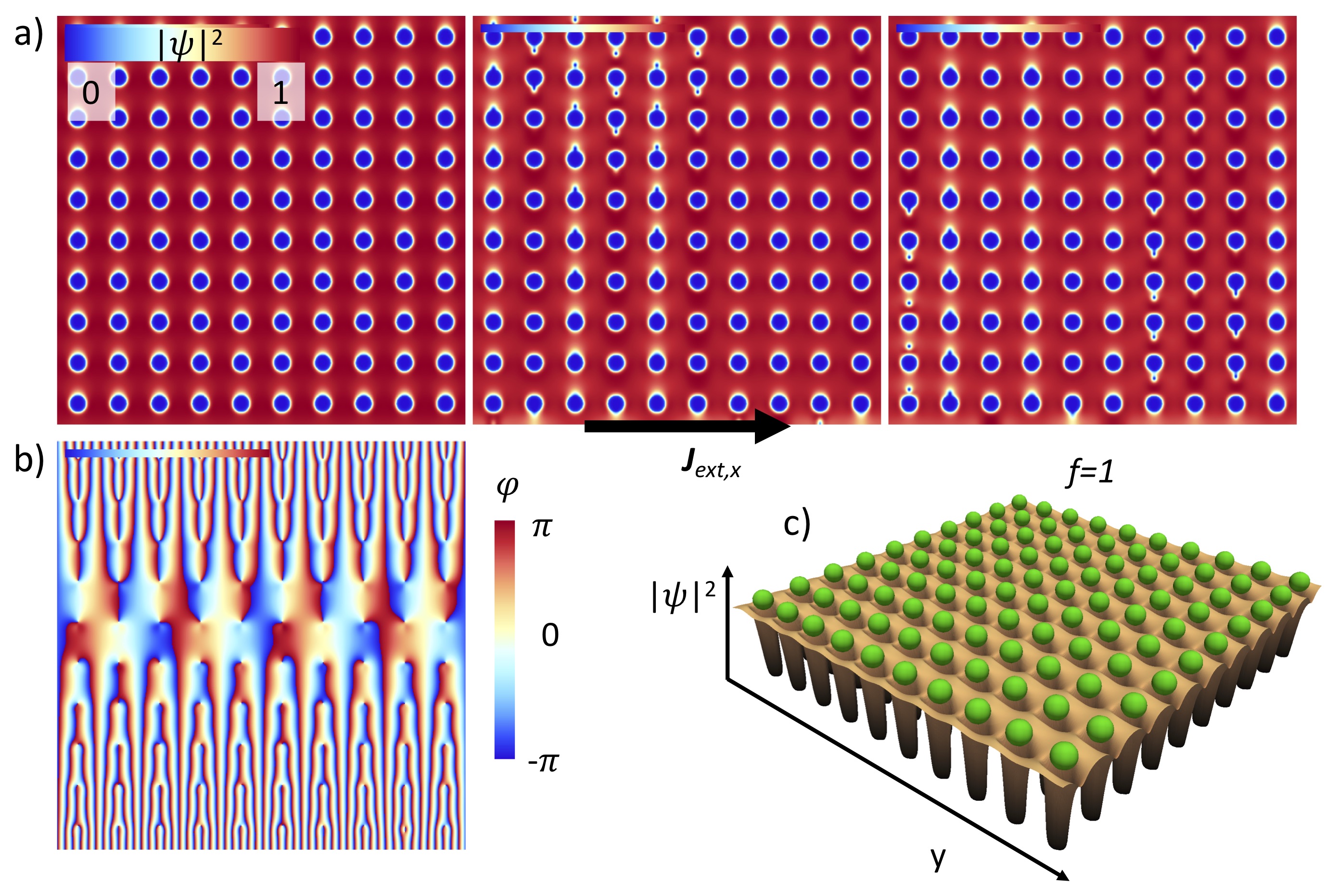}
	\caption{Simulation snapshots of the order parameter amplitude below the depinning transition in a) at the matching field ($f=1$), i.e., each inclusion is occupied by a single vortex. One sees how vortices start move (up) by creating avalanches in individual columns. b) shows the related phase of the order parameter [for first panel in a)], revealing the location of the vortices. A combined snapshot of vortex location (green spheres) and order parameter amplitude (surface plot) is shown in c). }
	\label{fig:depin}
\end{figure*}

\subsection{Modeling of the proximity pattern} 
The proximity array will be modeled as a superconducting film comprising a regular array of cylindrical metallic inclusions or holes.

To this effect we use so-called $\delta \Tc$ pinning, where the critical temperature is spatially modulated due to defects that cause pair-breaking scattering.\cite{Kwok:2016,Berdiyorov:2012a,Berdiyorov:2012b} We use the dimensionless coefficient of the linear term in Eq.~\eqref{eq:GL}, $\epsilon(\mathbf{r}) \propto \Tc(\mathbf{r}) - T$, to introduce spatial $\Tc$ modulations. This means that for $\Tc(\mathbf{r}) < T$, the linear coefficient is negative, which models normal and insulating defects, while for $\Tc(\mathbf{r}) > T$ and different from the bulk $\Tc$, weak superconducting defects are captured. We use this modulation to both model the inherent defects of the material as well as the larger-scale  periodic pinning array. Our equation is scaled such that $\epsilon(\mathbf{r}) = 1$ for the bulk superconductor (due to the choice of the length scale), i.e., $\epsilon(\mathbf{r}) = \ei = (\Tci - T) / (\Tc - T)$, where $\Tci$ is the critical temperature inside the inclusions and $\Tc$ is the bulk critical temperature. The pinning centers of the array are modeled by short cylinders of diameter~$d$ and height~$h$. Together with the value of $\ei$, these are the parameters controlling the pinning properties of the defects.

Our benchmark system has a size of $L_x\cdot L_y=(250\xi)^2$ with thickness $h=\xi$, and is discretized in $512^2$ grid points. The diameter of the inclusions is, if not stated otherwise, chosen as $d=8\xi$ and the lattice constant $\ell=10\xi$. Thus, such an array has $2\xi$ wide narrow weak links between neighboring inclusions forming effectively Josephson junctions (jj), see Fig.~\ref{fig:system}. The `strength' of the inclusion is chosen as $\epsilon=-10$. The whole system has periodic boundary conditions in $x$-direction and open/no-current boundary conditions in $y$-direction.
We characterize the density of inclusions by the matching field $B_{\phi} = \phi_0/\ell^2$. At $B = B_{\phi}$ the number of vortices is equal to the number of defects. The quantity $f=B/B_{\phi}$ is then the `filling' factor of the proximity array by vortices.
In dimensionless units, the matching field is given by $2\pi N_i/(L_xLy)$, where $N_i$ is the number of inclusions. Here $N_i=25^2$ and therefore $B_{\phi}\approx 0.0628$.

\subsection{Calculation of voltage/resistance and differential magnetoresistance}
Voltage, resistance, and differential resistance are obtained by averaging steady states for fixed external current and magnetic fields over many time steps and, if applicable, several intrinsic disorder realizations of the material.

For the evaluation of the differential magnetoresistance (MR) in the proximity array we use a special lock-in method to avoid strong fluctuations. 
Instead of fixing the current, we add an small rectangular AC oscillation to the current and average the voltage/resistance over each second half of each half-period before taking the difference. This is repeated and averaged several times for each fixed (DC) $\Jext$.

Here we investigate the MR for magnetic field close to $f=1$ and ramp the current from zero up to some value below the critical current. At the end of each cycle we average the voltage across the sample. 

\section{Results}

Using the approach described above, we investigated the dynamic behavior of the proximity arrays. First we studied the vortex dynamics close to the matching field and then studied the magnetoresistance to some detail to uncover the vortex Mott transition.

\subsection{Vortex dynamics}
\begin{figure}[htb]
	\includegraphics[width=\columnwidth]{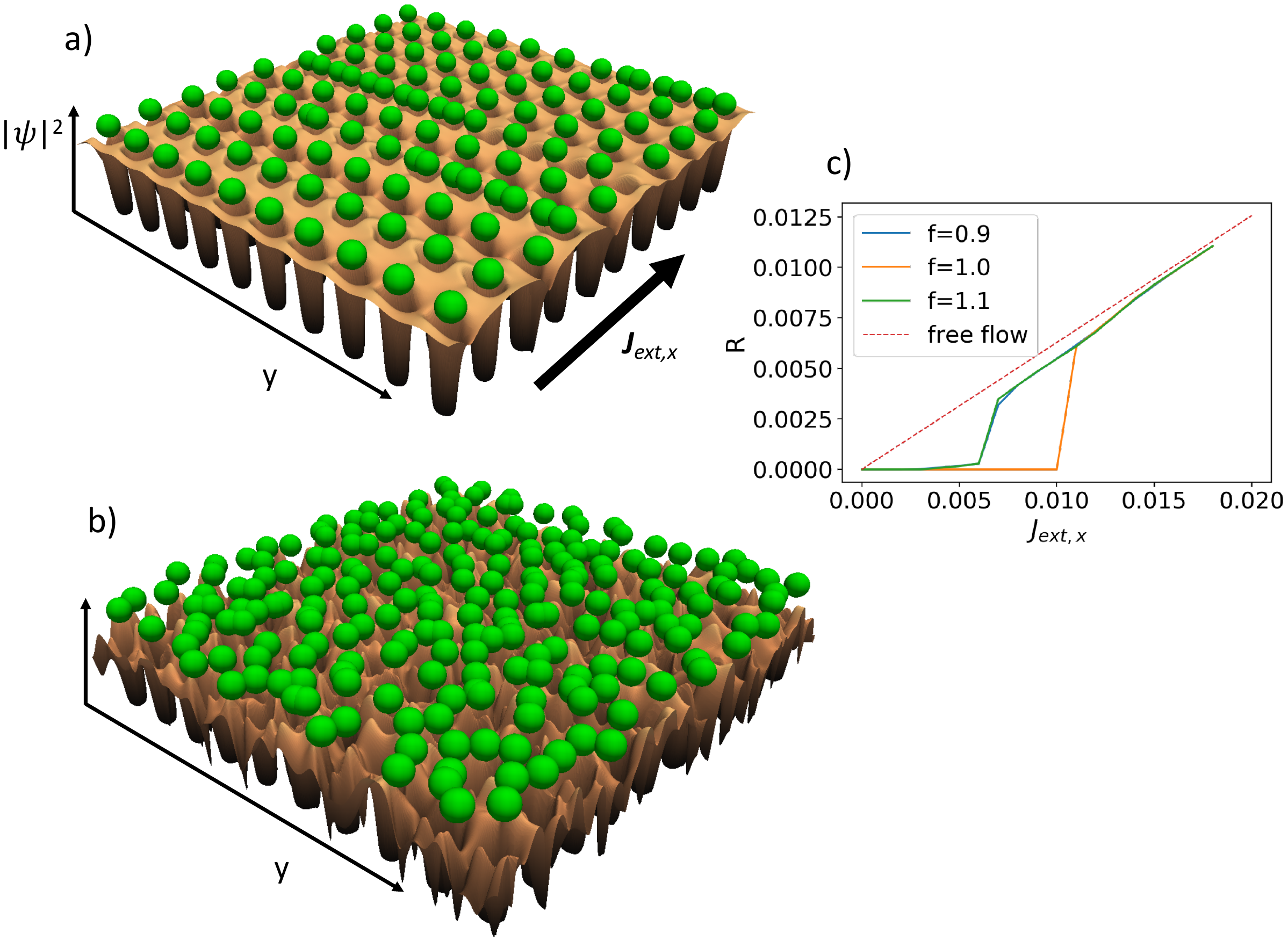}
	\caption{Vortex location and order parameter amplitude as in Fig.~\ref{fig:depin}c) at larger currents, illustrating the depinning process as columnar avalanches in a) and a highly dynamic state at larger currents in b). See also supplementary movie~\cite{movie} illustrating the vortex dynamics at the matching field. c) shows the I-V characteristics for $f=0.9,1.0,1.1$ at the critical current. For comparison, the depairing current in these units is of order $0.3$.}
	\label{fig:dyn}
\end{figure}

Below the critical current the system is in an insulating state and vortex motion is determined by creep and slow motion through the weak links, i.e. phase slips. Therefore, vortices at the matching field are mostly pinned in the inclusions. Figure~\ref{fig:depin} shows a typical motion event near the critical current (or depinning transition), when vortices randomly leave an inclusion, which causes an avalanche process in a column for the proximity array. Panel a) shows the amplitude of the complex order parameter and in b) a typical phase configuration is shown, which corresponds to the first plot in panel a) and illustrates the position of the vortices. Using a gauge invariant vortex detection method~\cite{Phillips2015}, we can extract the precise location of the vortices and superimpose their position with the order parameter configuration, shown in panel c).

Figure~\ref{fig:dyn} shows two additional vortex configuration plots at larger currents in panels a) and b). Panel c) shows the I-V characteristics for three different filling factors, which illustrates the matching effect in the proximity array, i.e., that the critical current  has a local maximum at $f=1$. An animation of the depinning dynamics for $f=1$ can be found in the supplementary movie~\cite{movie}.

These studies give a first hint on the existance of a vortex Mott transition in these systems. For a more detailed investigation, we need to study the magneto resistance and the differential magneto resistance.

\subsection{Magnetoresistance}
\begin{figure}[htb]
	\includegraphics[width=0.6\columnwidth]{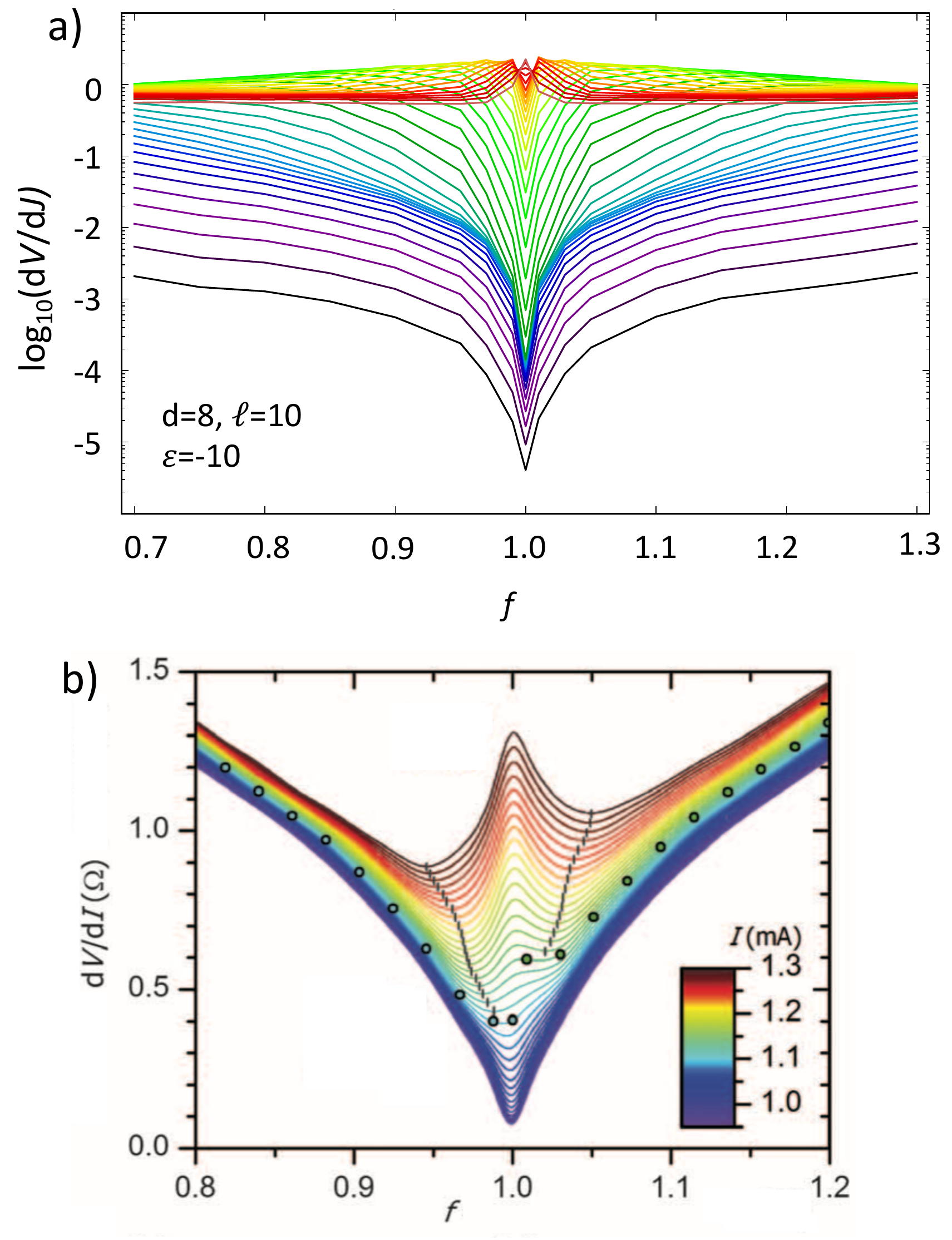}
	\caption{Current-assisted Mott transition. Differential conductance as functions of the filling factor $f = B/B_{\phi}$, where $B$ is the applied magnetic field and $B_{\phi}$ is the matching field characterizing inclusion pattern. Black color corresponds to zero applied critical current and red corresponds to approximately 60–80\% of the critical current obtained by simulations for $d=8\xi$ and $\ell=10\xi$ and insulating inclusions with $\epsilon=-10$ (see text). In b) we show an experimental measurement for a Nb island array for comparison, see~\cite{Science:2015}.}
	\label{fig:diffMR}
\end{figure}

The  result for differential resistance $R_d = dV/dJ$ is presented in Fig.~\ref{fig:diffMR}a) as a function of the filling factor $f = B/B_{\phi}$ for different currents. 
At smallest currents, both $V(f)\sim R(f)$ and $R_d(f)$ displays dips near $f = 1$ as would naively expect from the matching effect: All vortices are pinning perfectly by the inclusion. Deviations from the matching field lead to either excess vortices or vortex vacancies on the inclusion positions.
However, upon increasing current, the dip of $R_d$ near $f = 1$ turns into a peak, marking the current-driven Mott transition. 
This is clearly reproduced by the simulations for different magnetoresistance curves at different fixed applied currents $\Jext$, shown in Fig.~\ref{fig:diffMR}a). This behavior is very similar to the experimentally observed differential resistance reversal in arrays of superconducting Nb islands~\cite{Science:2015} and shown in panel b).

\begin{figure}[htb]
	\includegraphics[width=\columnwidth]{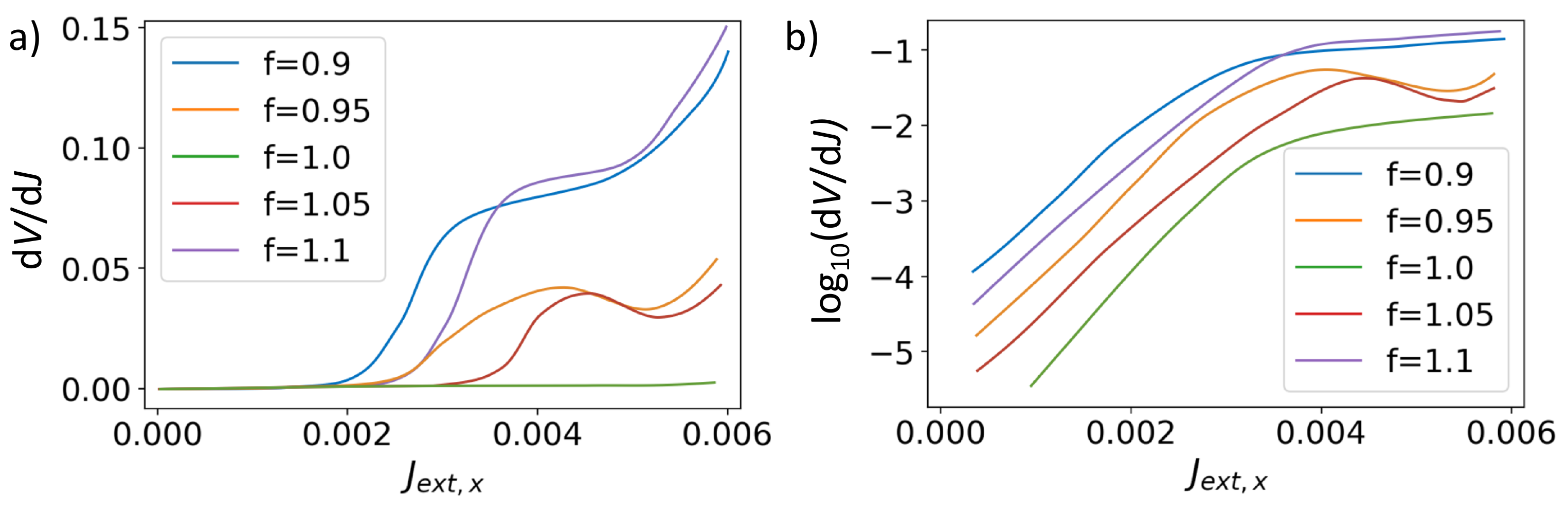}
	\caption{Differential resistance at constant filling factors $f$. Close to $f=1$ the differential resistance shows non-monotonic behavior. a) on linear scale and b) on semi-logarithmic scale.}
	\label{fig:dVdI}
\end{figure}

However, there is an important difference: the differential resistance shows non-monotonic behavior near $f=1$, which can be seen by plotting the current dependence at fixed magnetic fields, which is shown in Fig.~\ref{fig:dVdI}.
In this respect the obtained results differ also from those observed in recent experiments on the similar proximity array\,\cite{Marcus2018}, where the metallic phase was associated with an anomalous metal. Our results indicate that the metallic phase in the array of inclusions may have somewhat different properties that call for further investigation.

To conclude, our studies reveal that the vortex Mott transition can be observed in two-dimensional proximity arrays by performing TDGL simulations. However, in contrast to the experimentally observed monotonic behavior of the differential MR in superconducting island arrays, where the islands are separated by a substrate, our case shows non-monotonic behavior.
As a result, the scaling behavior of the experiment cannot be applied to our proximity arrays. The difference could be explained by the different vortex dynamics: In the proximity array vortices have to cross the weak links via phase-slip processes~\cite{Varlamov2015,Ovchinnikov2020}, while in the island array vortices do not need to suppress a residual superconducting order parameter. Furthermore, at $f=1$ the vortex system depinns collectively as shown in Fig.~\ref{fig:dyn}c), while in the vacancy or excess situation allows for individual vortex motion.

\begin{acknowledgments}
This work was supported by the U.S. Department of Energy, Office of Science, Basic Energy Sciences, Materials Sciences and Engineering Division. The computational work is done on the GPU cluster in the Materials Science Division at Argonne National Laboratory.
\end{acknowledgments}

\end{document}